\pdfoutput=1
\documentclass[fleqn,usenatbib]{mnras}

\usepackage{newtxtext,newtxmath}

\usepackage[T1]{fontenc}

\DeclareRobustCommand{\VAN}[3]{#2}
\let\VANthebibliography\thebibliography
\def\thebibliography{\DeclareRobustCommand{\VAN}[3]{##3}\VANthebibliography}

\usepackage{graphicx}	
\usepackage{amsmath}	

\title[Simulations of expanding SN shells near SMBHs]{\textsc{flash}-light on the \textsc{ring}: hydrodynamic  simulations of expanding supernova shells near supermassive black holes}

\author[B. Barna et al.]{
B. Barna,$^{1,2}$\thanks{E-mail: bbarna@titan.physx.u-szeged.hu}
J. Palou\v {s},$^{1}$
S. Ehlerov\' a,$^{1}$
R. W\" unsch,$^{1}$
M. R. Morris,$^{3}$
P. Vermot$^{1}$
\\
$^{1}$Astronomical Institute, Academy of Sciences, Bo\v{c}n\'{\i} II 1401, Prague, Czech Republic\\
$^{2}$Physics Institute, University of Szeged, D\'{o}m t\'{e}r 9, Szeged, 6723, Hungary\\
$^{3}$Department of Physics and Astronomy, University of California, Los Angeles, CA 90095-1547, USA
}

\date{Accepted XXX. Received YYY; in original form ZZZ}

\pubyear{2015}

\volume{{\rm in press}}
\begin{document}
\label{firstpage}
\pagerange{\pageref{firstpage}--\pageref{lastpage}}
\maketitle

\begin{abstract}
    The way supermassive black holes (SMBH) in galactic centers accumulate their mass is not completely determined. At large scales, it is governed by galactic encounters, mass inflows connected to spirals arms and bars, or due to expanding shells from supernova (SN) explosions in the central parts of galaxies. The investigation of the latter process requires an extensive set of gas dynamical simulations to explore the muti-dimensional parameter space needed to frame the phenomenon. The aims of this paper are to extend our investigation of the importance of supernovae for inducing accretion onto a SMBH and carry out a comparison between the fully hydrodynamic code Flash and the much less computationally intensive code Ring, which uses the thin shell approximation. 
We simulate 3D expanding shells in a gravitational potential similar to that of the Galactic Center with a variety of homogeneous and turbulent environments. In homogeneous media, we find convincing agreement between \textsc{flash} and \textsc{ring} in the shapes of shells and their equivalent radii throughout their whole evolution until they become subsonic. In highly inhomogeneous, turbulent media, there is also a good agreement of shapes and sizes of shells, and of the times of their first contact with the central 1 pc sphere, where we assume that they join the accretion flow. The comparison supports the proposition that a SN occurring at a galactocentric distance of 5 pc typically drives  1 - 3 $M_\odot$  into the central 1 pc around the galactic center. 
\end{abstract}

\begin{keywords}
supernovae: general -- ISM: supernova remnants -- Galaxy: centre -- galaxies: active
\end{keywords}



\section{Introduction}

Encounters between galaxies cause perturbations of their gravitational potentials that can open the door for mass inflow towards the nuclei of those galaxies. Spiral arms and galactic central bars can be other reasons for angular momentum redistribution and a growing central mass concentration. However, despite many decades of intensive research, the relative importance of various contributing mechanisms has still not been determined. One of the possible processes is the accretion of clouds, which could be driven by nearby SN explosions, gaining mass from the interstellar matter (ISM) and eventually injecting some matter into the vicinity of the SMBH. We explored this possibility in a previous paper by \citet{Palous20}[Paper I]. Here we extend our investigation to lower densities and inhomogeneous turbulent media.   

An expanding SN shell goes through different evolutionary stages \citep[for a detailed description see e.g.][]{Ostriker88,Bisnovatyi-Kogan95}. Very soon after the explosion, after $t_\rmn{exp} \approx 100$ s, the expansion becomes homologous and the ejecta propagates without any significant kinetic losses. In this free-expansion phase, the gas of the ambient medium is compressed and swept up by the shock. 
As the mass of the swept-up gas becomes comparable to the mass of the ejecta, a reverse shock propagating backward to the explosion site is formed. During this so-called Sedov-Taylor phase, the radiative losses are negligible and the expansion is adiabatic. The hot inner gas drives further expansion of the forward shock. Later, as the density due to mass accumulation increases, the shock front slows down, and its temperature drops since the thermal energy of the expanding shell is radiated away. At this time a thin shell is formed. Even later, when the internal pressure drops to the level of the external pressure, the thin shell is no longer pushed from inside of the bubble, and it enters the so-called "snowplow" momentum conservation phase. Finally, as the expansion velocity drops below the local sound speed of the ambient medium, the SN remnant disperses into the ISM. Inside several tens of pc from the galactic center, the evolution of an expanding SN shell is typically  completed within a few tens of thousands to a few hundreds of thousands of years.

Following a given SN explosion, the evolution of its expanding remnant depends on many parameters that may influence shape, velocity, and total mass collected in the expanding shell. An analysis of the full parameter space with a full hydrodynamic treatment would consequently be rather expensive. 
In this long-term project, for which we aim to perform many computations, it is very important to use the most parsimonious code possible.
Our intent  is to explore the parameter space with the fast code \textsc{ring} using the thin-shell approximation. To test its validity we compare results to a fully hydrodynamic code, \textsc{flash}, at selected location of the SN explosion in both homogeneous and inhomogeneous media. This comparison also serves the community because it shows that \textsc{ring} can be used as an alternative to the popular but much more computationally intensive code \textsc{flash} for investigation of the dynamical effects of expanding supernova shells in complex environments.

The paper is organized as follows: we first introduce the simulation domain for the homogeneous ISM distribution near the galactic center (GC), by describing the physical and computational characteristics in Sec. \ref{setup}. We then introduce the results from the simulation with a uniform ambient medium in Sec. \ref{uniform}.  The turbulent medium simulations and the \textsc{flash} versus \textsc{ring} comparison are described in Sec. \ref{turbulent}. 
Finally, we summarize our results in Sec. \ref{conclusions}. 

\section{Simulation setups for \textsc{ring} and \textsc{flash}}
\label{setup}

Supernova remnants near a SMBH were studied by \cite{Palous20} with the code \textsc{ring} applied to the case of shells expanding into a homogeneous distribution of the ISM and subjected to the gravitational potential of the SMBH and of the \textbf{nuclear star clusters (NSCs)} of the Milky Way. They compared the results of \textsc{ring} simulations with a one-dimensional version of full hydrodynamic simulations using the code \textsc{flash},  during early stages of expansion up to 3 kyr after the SN explosion. Here, we use the 3D versions of both codes and extend the comparison to longer expansion times,  when parts of the shell have slowed to subsonic velocities and the shell starts to dissolve.       

\subsection{Gravitational potential and rotation}
\label{sec:grav_rot}
We use the same form of the gravitational field and of the rotation as in \citet{Palous20}, which fits the mass distribution at the center of the Milky Way. Two components of the NSC can be distinguished there, and the rotational velocity field is set according to the total gravitational potential of the SMBH and the NSC.  The implied density distribution is given by \citet{Chatzopoulos15}, which can be described with the two component $\gamma$ model \citep{Dehnen93}:

\begin{equation}
    \rho_\rmn{NSC} (r_\rmn{GC}) = \sum_{i=1}^{2} \frac{3 - \gamma_i}{4\pi} \frac{M_\rmn{NSC_i} a_i}{r_\rmn{GC}^{\gamma_i} (r_\rmn{GC} + a_i)^{4-\gamma_i}}
\end{equation}
where $M_\rmn{NSC_i}$ is the mass and $a_i$ is the characteristic radius of the individual stellar subclusters, and $r_{GC}$ is the galactocentric distance. Here we adopt according to \citet{Chatzopoulos15} the following set of values: $M_\rmn{NSC_1} = 2.7 \times 10^7$ M$_\odot$, $a_1 = 3.9$ pc, $\gamma_1 = 0.51$; $M_\rmn{NSC_2} = 2.8 \times 10^9$ M$_\odot$, $a_2 = 94.4$ pc, $\gamma_2 = 0.07$.

The point-like supermassive black hole and the extended mass distribution of the NSCs result in the following gravitational potential \citep{Chatzopoulos15}:
\begin{equation}
\Psi (r_\rmn{GC}) = -\frac{G M_\rmn{SMBH}}{(r_\rmn{GC} + \epsilon)} - \sum_{i=1}^{2}\frac{G M_\rmn{NSC_1}}{a_\rmn{i}(2 - \gamma_\rmn{i})} \Bigg(1 - \bigg(\frac{r_\rmn{GC}}{r_\rmn{GC} + a_\rmn{i}}\bigg)^{2-\gamma_\rmn{i}}\Bigg),
	\label{eq:potential}
\end{equation} 
where $M_\rmn{SMBH} = 4 \times 10^6 M_\odot$ is the mass of the SMBH, and $\epsilon$ is a small constant equal to $\sim 10^{-6}$ pc, inserted to prevent extremely large values of the potential and of its derivatives. 

Gas in the ambient medium in the gravitational field of the SMBH and NSC rotates with the velocity:
\begin{equation}
v_\rmn{rot}(R,z) = 
\left( \frac{d\Psi (r_\rmn{GC})}{dr_\rmn{GC}}r_\rmn{GC} \right)^{1/2} \, \frac{R}{r_\rmn{GC}}
	\label{eq:quadratic}
\end{equation} 
where $(R, z)$ are the galactocentric cylindrical coordinates.  The galactocentric distance is  $r_\rmn{GC} = \sqrt{R^2 + z^2}$.

\subsection{Initial conditions}

\cite{Palous20} discuss the importance of the SN position relative to the SMBH for mass delivery into the central parsec of the galaxy in the case of a homogeneous ISM distribution within the central region of the galaxy. There are many other factors that may influence the evolution of an SNR, such as the 
the mass of the SMBH and NSC, the density distribution of ISM, inflows and outflows from the nuclear star cluster (NSC), the presence of a central accretion disk, etc. These factors will be discussed in separate communications. Here we focus on the position of the SN explosion relative to the GC and on the importance of the structure of the ISM.

We place the SN explosion at different locations within the distance $r_{GC} = 20$ pc from the GC, 
inserting ejecta with a total mass of 10 $M_\odot$ and a kinetic energy of 10$^{51}$ erg. 
The results of \textsc{flash} and \textsc{ring} simulations are compared for supernovae occurring at several selected locations.
In the case of \textsc{flash} the ejecta is inserted into a spherical volume $(r_{shell} \sim 0.25 pc)$ with uniform properties and an initial expansion velocity of $10^4$ km/s. \textsc{ring} simulations start later, at the time of shell formation, and its total initial mass is that of the ejecta plus the mass collected from the ISM during early expansion with kinetic and thermal energies taken from \textsc{flash} simulations, as given in Table \ref{tab:initial_parameters}. This early evolution of the SN remnant is discussed by \cite{Palous20}.




\subsection{Simulation setup for FLASH}
\label{sec:setup_flash}
 \textsc{flash} is the grid-based adaptive mesh refinement hydrodynamic code  \citep{Fryxell00} using the Piecewise Parabolic Method \citep{Colella84} with the time-step controlled by the Courant-Friedrichs-Lewy criterion. It uses a grid structure of blocks as the computational domain. A root block, which corresponds to refinement level 1, is split into eight sub-blocks, each with equivalent volume (refinement level 2). The subdivision of blocks continues until the blocks can effectively resolve the process of interest in the computational domain. However, to avoid extremely high computational costs in our 3D simulations, we set the smallest size of a grid cell to 0.028 pc within a sphere of radius 10 pc around the galactic center affected by propagation of the shell (12 pc for the low ambient medium density case). The regions beyond the 10 pc (not affected by the evolving SN shell) are treated with eight times larger grid cell width.

Radiative losses are calculated following the prescription of \citet{Schure09}, assuming solar metallicity. As a test of the accurately functioning cooling, the thin shell is formed at the same time, as in the 1D simulations presented in \cite{Palous20} (see Tab. \ref{tab:initial_parameters}), which also provide initial parameters for our RING simulations.

We do not attempt to describe dynamical processes occurring very close to the galactic center, where the circumnuclear  disk of the SMBH dominates and where outflows can change the motion of the ISM. Also, in order to avoid extremely high rotational velocities in the inner parts of our simulation closest to the galactic center, the rotational curve is changed: within 0.5 pc of the galactic center, we adopt linearly increasing rotation with $R$  in the $z = 0$ plane, with the decrease of it proportional $R/r_{GC}$ for increasing $z$. Note that this change does not affect the evolution of the shell during the studied time period outside $r_{GC }$ = 0.5 pc.


Our challenge is to perform a comparison of \textsc{flash} to \textsc{ring} simulations, and for that we need to define in \textsc{flash} which cells should be considered as a part of the thin shell and which are outside, so that we can derive the total shell mass ($M_\rmn{shell}$) and its kinetic energy ($E_\rmn{kin}$). 
For this purpose we adopt a criterion related to the definition of the shock front, which propagates with supersonic velocity. While the ambient medium has Mach number close to zero, the cells of the shell behind the shock have high Mach numbers. We define a local Mach number, 
\begin{equation}
  M_\rmn{FLASH} = \frac{\vec{v} - \vec{v_\rmn{rot}}}{(1.15\, \, km s^{-1})},   
\end{equation}

\textbf{where $\vec{v}$ and $\vec{v_\rmn{rot}}$ are the gas velocity and the orbital velocity given by Eq.~(3), respectively, while their difference is divided by the sound speed in the ambient medium with $T_\rmn{ISM} = 10^2$\,K.} To precisely distinguish the supernova bubble from the ambient medium around it, only cells with Mach number $M_\rmn{FLASH} > 3$ are assumed to be part of the shell. To follow these mass components, the field parameter S$_\rmn{sh}$ is introduced, which is initialized to zero for all cells. Once a cell fulfills the Mach number criterion, its mass is marked with S$_\rmn{sh}=1$. Subsequently, the $S_\rmn{s}$ field is advected by the hydro solver, and therefore the gas, once marked as a part of the shell, will retain this marking ever after. At the edge of the shell, the $0 < S_{sh} < 1$ values appear due to the mixing of the shell with the ambient gas during the shock propagation. To compute the total mass of the shell, we consider those cells with S$_\rmn{sh} > 0.5$, i.e., where the mass belonging to the shell represents the majority of the cell mass.


\begin{table}
\centering
\begin{tabular}{cccccc}
$n_\rmn{ISM}$ & $t_\rmn{shell}$ & $r_\rmn{shell}$ & $m_\rmn{shell}$  & $E_\rmn{th} / E_\rmn{tot}$ & $E_\rmn{kin} / E_\rmn{tot}$\\
  cm$^{-3}$ & yr    & pc   & $M_\odot$ &  &    \\
\hline
10$^5$   &   150   &   0.3   &   180   &   0.23   &   0.21\\
10$^4$   &   500   &   0.7   &   210   &   0.24   &   0.27\\
10$^3$   &   2000   &   1.8   &   400   &   0.23   &   0.28\\
10$^2$   &   5700   &   4.5   &   550   &   0.29   &   0.32\\
\hline
\end{tabular}
\caption{Initial parameters of the SN shells derived from 1D \textsc{flash} simulations. 
}
\label{tab:initial_parameters} 
\end{table}


\begin{figure*}
    \centering
	\includegraphics[width=13cm]{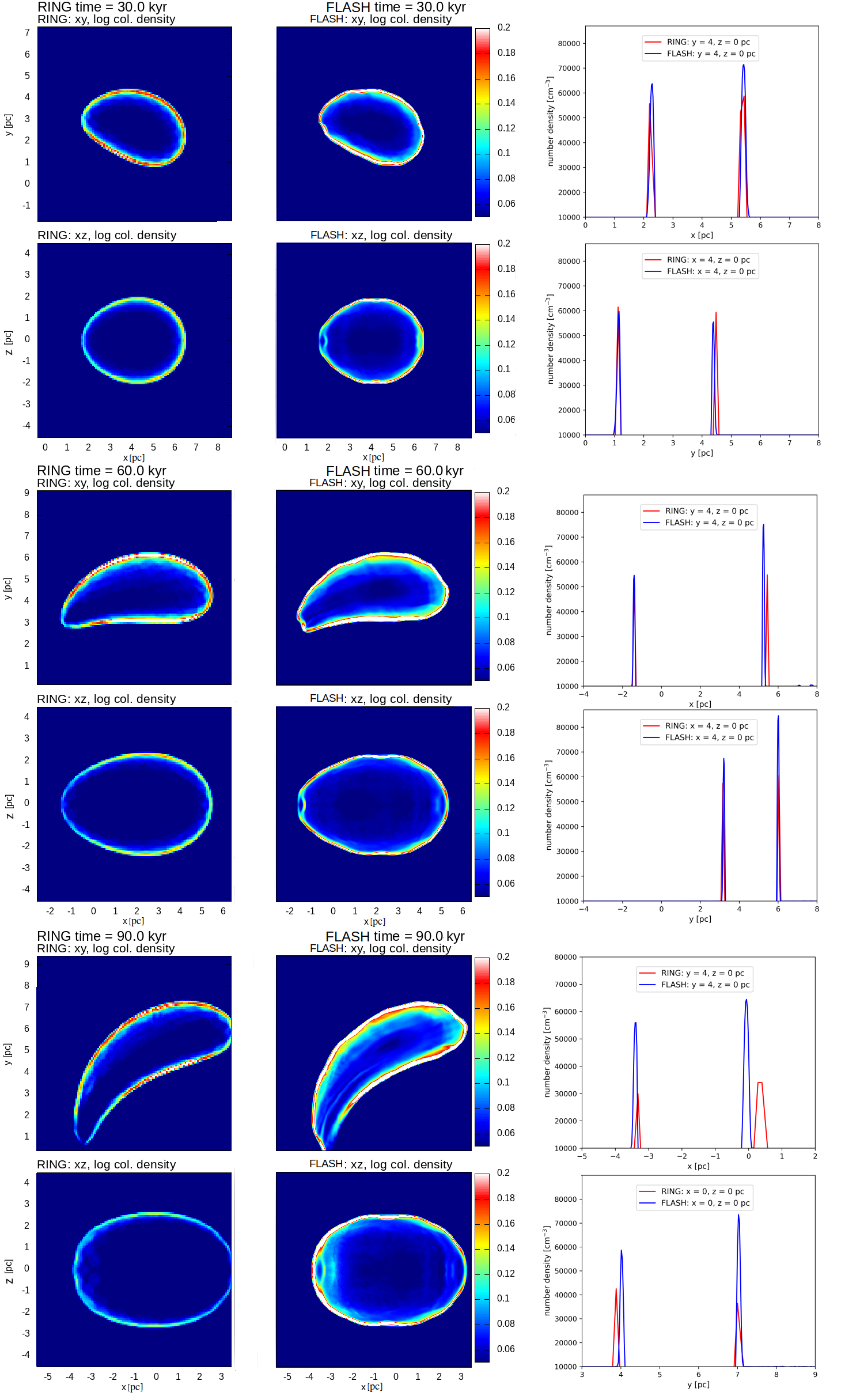}
    \caption{The evolution of the SN shell in \textsc{ring} and \textsc{flash} simulations at three different times in the N4 model of expansion starting at (x, y, z) = (5, 0, 0) pc. The left and center columns show the column density in g cm$^{-2}$ projected onto the (x, y) and (x, z) planes for \textsc{ring} and \textsc{flash}, respectively. \textbf{The right column shows the one-dimensional cuts of number density at the same time steps; in the case of \textsc{ring}, the infinitesimally thin shell is resampled with resolution of 0.12 pc.}
    }
    \label{fig:sn_shell_evolution}
\end{figure*}


\begin{figure}
    \centering
	\includegraphics[width=\columnwidth]{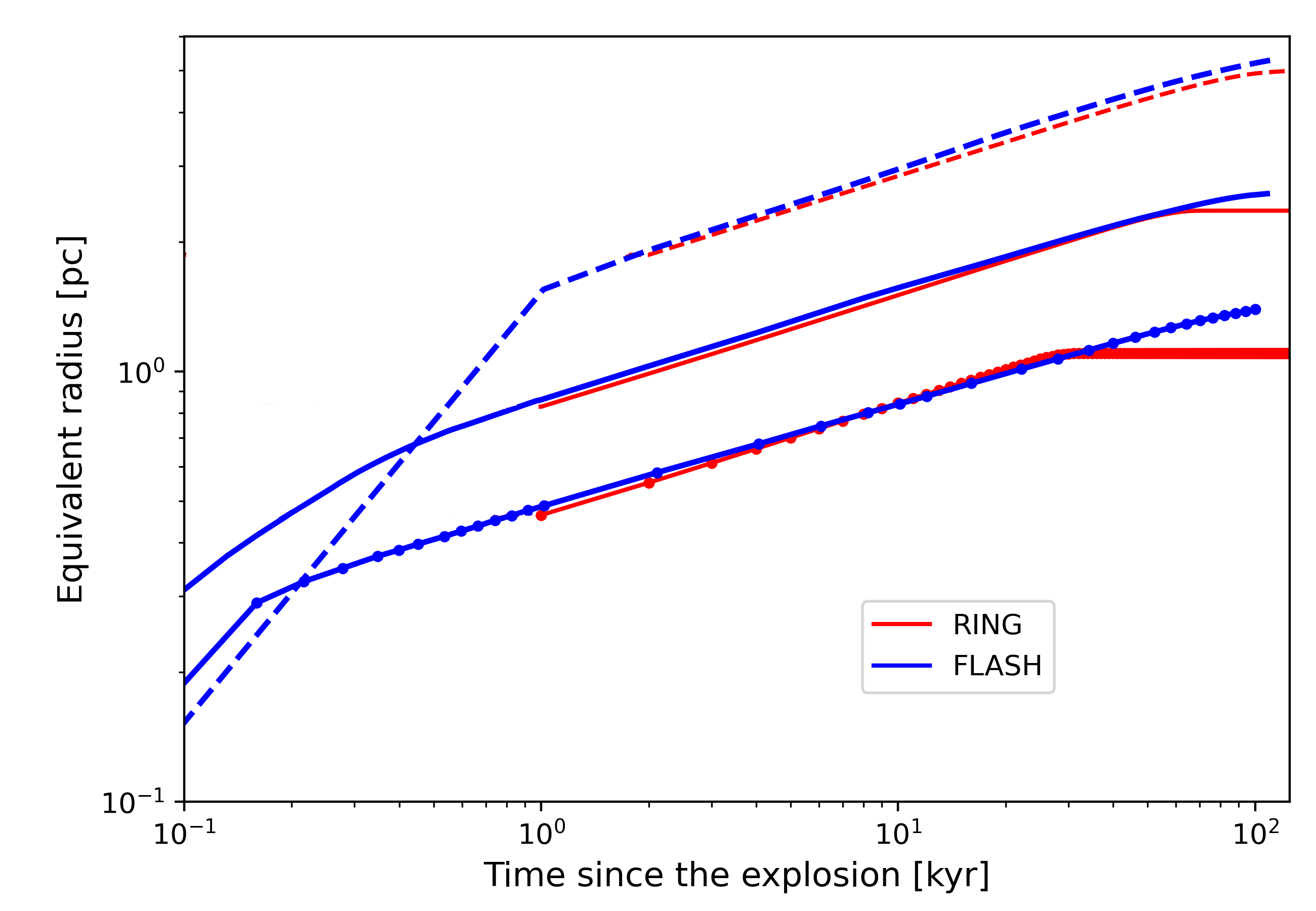}
    \caption{The equivalent radius of the SN shell as a function of time in the \textsc{ring} (red) and \textsc{flash} (blue) simulations, assuming the ambient medium number density of $n_\rmn{ISM}=10^3$ (dashed), 10$^4$ (solid) and 10$^5$ (solid with dots) cm$^{-3}$. 
    }
    \label{fig:sn_shell_radius-N3-N4-N5}
\end{figure}

\begin{figure*}
    \centering
	\includegraphics[width=13cm]{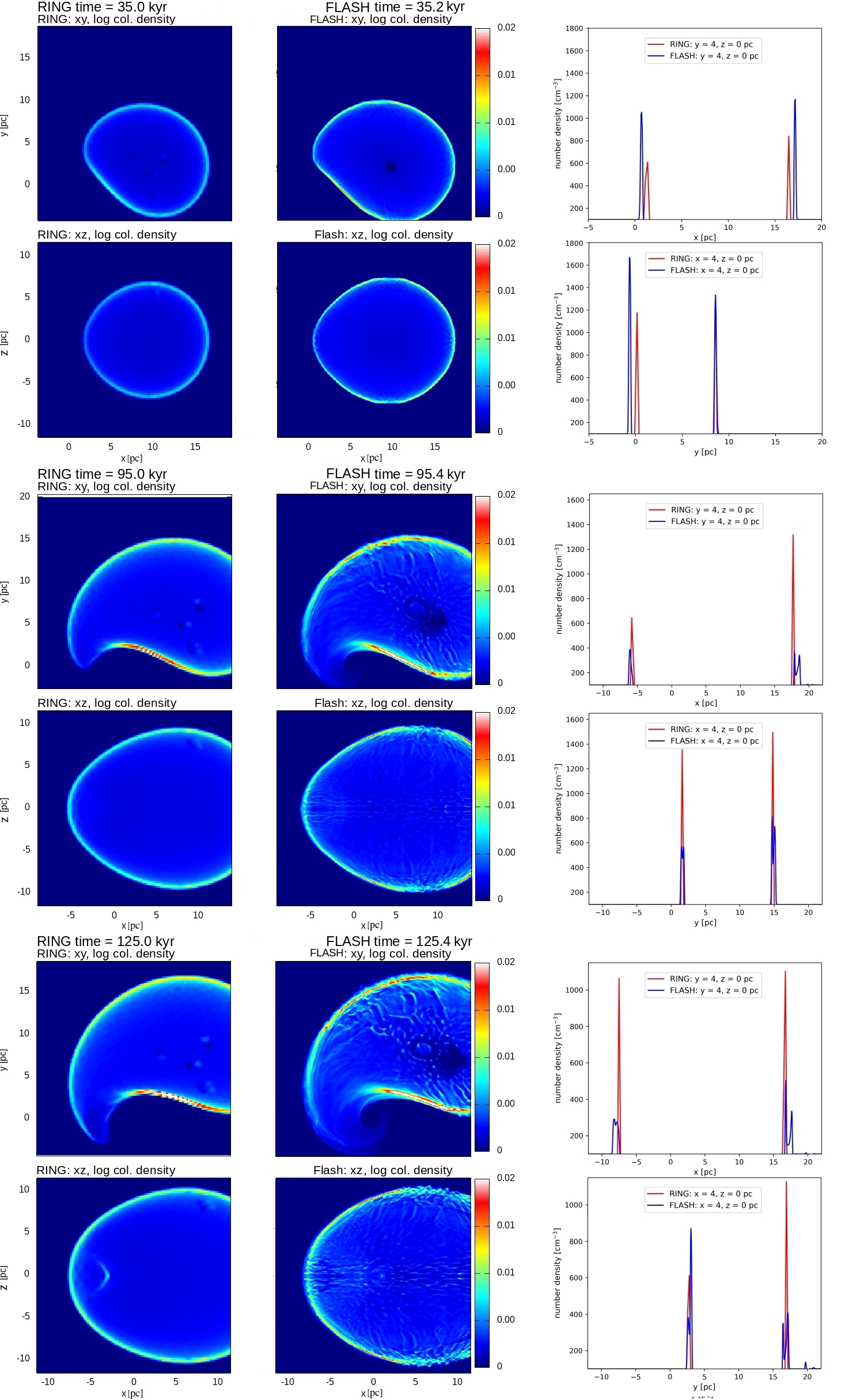}
    \caption{The evolution of the SN shell in \textsc{ring} and \textsc{flash} simulations at three different time steps in the N2 model of expansion starting at (x, y, z) = (10, 0, 0) pc. The left and center columns show the column density in g cm$^{-2}$  projected onto the (x, y) and (x, z) planes for \textsc{ring} and \textsc{flash}, respectively. \textbf{The right column shows the one-dimensional cuts of number density at the same time steps; in the case of \textsc{ring}, the infinitesimally thin shell is resampled with resolution of 0.24 pc.}
    }
    \label{fig:sn_shell_evolution_n2}
\end{figure*}

\begin{figure}
    \centering
	\includegraphics[width=\columnwidth]{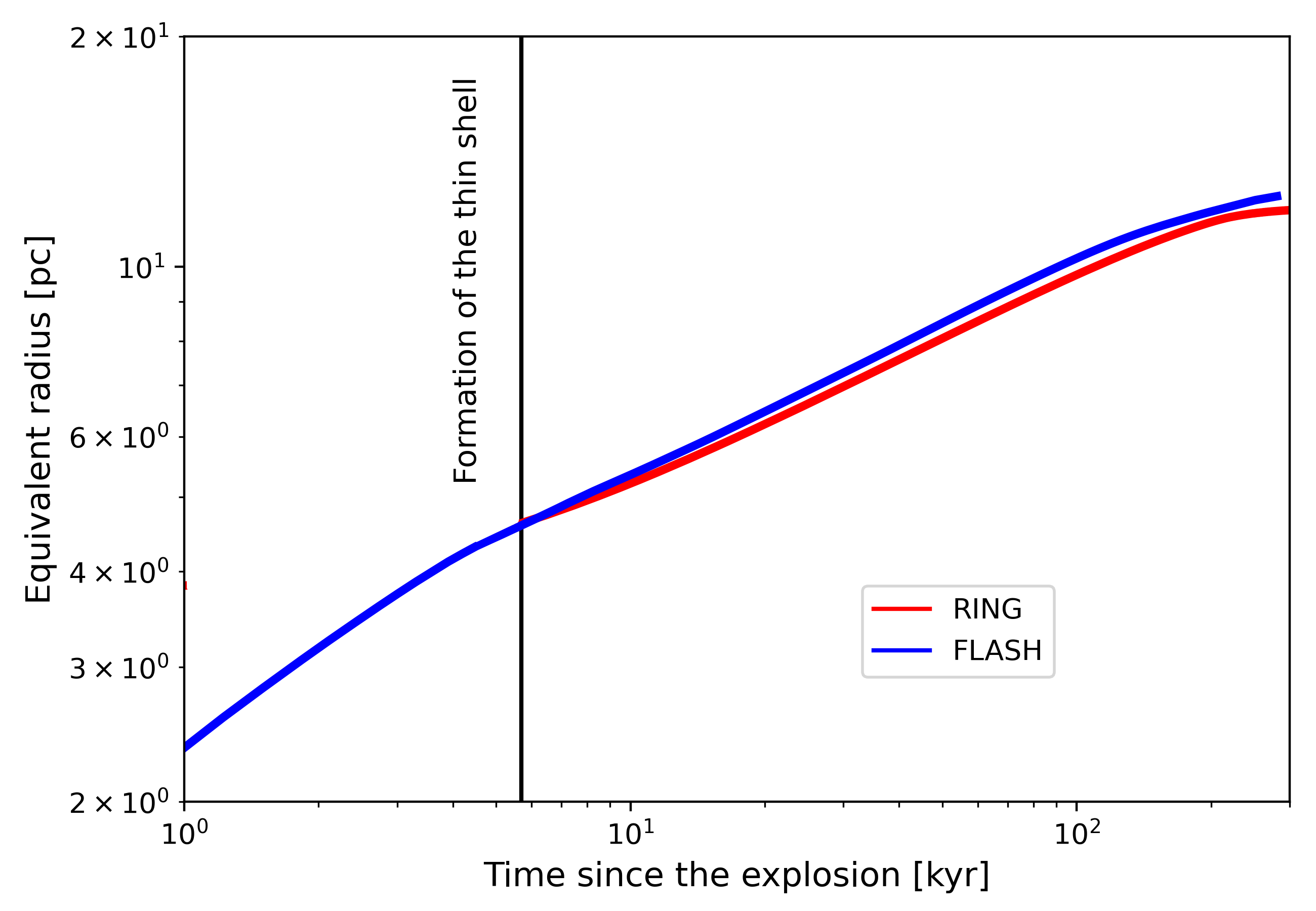}
    \caption{The equivalent radius of the SN shell as a function of time in \textsc{ring} (red) and \textsc{flash} (blue) simulations, assuming  ambient medium number densities of $n_\rmn{ISM}=10^2$ cm$^{-3}$.
    }
    \label{fig:sn_shell_radius-N2}
\end{figure}

\subsection{Simulation setup for RING}
\label{sec:ring}
RING is a 3D simplified hydrodynamic code for modelling the evolution and propagation of expanding gaseous shells. The main assumption is the thin-shell approximation, i.e. the assumption that the width of the wall of the shell is negligible compared to its diameter. The thin shell is divided into $N_\rmn{elem}$ elements, which are distributed equally on the initial sphere of the shell. As the thin shell rapidly expands, it accumulates mass from the ISM and eventually slows down. \textsc{ring} computes the trajectories of the shell elements in an external gravitational potential, by solving the equations of momentum conservation:

\begin{equation}
\frac{d}{dt}(m_i \vec{v_i}) + \rho_\rmn{out} \cdot [\vec{S_i} \cdot (\vec{v_i} - \vec{v_\rmn{out}})] \cdot \vec{v_\rmn{out}} = (P_\rmn{int} - P_\rmn{out}) \cdot \vec{S_i} + m_i \vec{g}
\end{equation}

\noindent
where $m$, $S$ and $v$ are the mass, the surface area vector and the expansion velocity vector of the i-th surface element, respectively.  $P_\rmn{out}$ and $\rho_\rmn{out}$ are the pressure and the density of the ISM outside of the shell, while $g$ is the local gravitational acceleration. The thin-shell approximation also assumes that the shell is dense enough to radiate all thermal energy produced by the shock compressing the ISM.

Each of the shell elements accretes the encountered ISM if their relative motion is supersonic. 

\begin{equation}
\frac{d}{dt}{m_i} = \rho_\rmn{out} \cdot [\vec{S_i} \cdot (\vec{v_i} - \vec{v_\rmn{out}})]
\label{eq:accr}
\end{equation}

\noindent
The pressure behind the shell can be described as:

\begin{equation}
P_\rmn{int} = \frac{2 E_\rmn{th}}{3 V_\rmn{int}}
\end{equation}

\noindent
By disregarding the radiative cooling of the cavity behind the shell, the change of $E_\rmn{th}$ can be described with the adiabatic energy balance equation:

\begin{equation}
\frac{d E_\rmn{th}}{d t} = - \frac{d V_\rmn{int}}{d t} \cdot P_\rmn{int}
\label{eq:enercav}
\end{equation}

\noindent
where $P_\rmn{int}$ and $V_\rmn{int}$ are the pressure and the volume of thee inside cavity.

Note that the equations above describe a far more simple approach than a hydrodynamic code. 
While the hydrodynamic code solves the continuity equation, the equation of motion and the energy equation on a 3D grid consisting typically of tens to hundreds millions computational elements (for the cases described below), the \textsc{ring} code solves the equation of motion for only several hundred surfaces. The continuity equation is simplified taking into account only the accretion (equation \ref{eq:accr}), and the energy equation is solved for the gas in the cavity assuming its uniformity (equation \ref{eq:enercav}), i.e. for a single computational element only. For these reasons, \textsc{ring} requires only minutes on a typical PC, compared to the thousands of CPU hours required by a complex hydrodynamic simulation. The price for this performance gain is that \textsc{ring} can work only under the special assumptions described above. Evaluating whether the assumptions are valid for typical SN shells in the Galactic center and whether \textsc{ring} works as expected is one of the main aims of this work.

The theoretical background of this method was developed by \cite{Kompaneets60} and \citep{Blinnikov82} and used to match with observational data by e.g. \cite{Ehlerova96}, \cite{Silich96}, \cite{Ehlerova97}, \cite{Elmegreen02}, \cite{Ehlerova18}, etc. The detailed description of the code \textsc{ring} is given by \cite{Palous20}.

\section{Uniform ambient medium}
\label{uniform}

The ISM moves in the external gravitational field of the SMBH and NSC.
Since the z-component of the gravitational acceleration, \ -- i.e., the component perpendicular to the rotational plane -- is not balanced by rotation, the ISM starts to collapse, forming, during a relatively short time after the start of the simulation ($\sim$50 kyr), a flat disk around the $z=0$ plane. This is observed in the case of \textsc{flash} simulations. In the case of the \textsc{ring} simulations the ISM ambient to the expanding shell does not move in the $z$ direction and its density is kept constant. For the purpose of the comparison of the two codes, to prevent the collapse of the ambient medium towards the galactic plane, we discarded the z-component of gravity in both codes. Doing this enables a direct match between \textsc{ring} and \textsc{flash} over the whole computational volume. Note that this choice does not result in a realistic shell evolution in 3D, however it allows a comparison of the shape of the shell and its projected densities in 2D near the plane of symmetry.

\subsection{Densities $n_\rmn{ISM} = 10^3, 10^4$ and $10^5$ cm$^{-3}$}

First, we explore the simulations presented by \cite{Palous20}, where the ISM is assumed to be molecular hydrogen with adiabatic index of $\gamma = 1.667$ homogeneously distributed within the central region of the galaxy. The initial assumed number densities are $n_\rmn{ISM} = 10^3, 10^4$ and $10^5$ cm$^{-3}$, hereafter labeled as the N3, N4 and N5 models, respectively. 
Fig. \ref{fig:sn_shell_evolution} shows the comparison of column densities projected onto the x-y and x-z planes at three different epochs in the case of N4 models. 

The surface of the shell is relatively smooth through the entire time period of the simulation. The commonly observed Rayleigh-Taylor instability is not present in our simulations, because it occurs in the very early phase of the SN remnant when the shell mass is dominated by the ejecta \citep{Martinez18}. Our approach simulates SN remnants in later phases when the shell is dominated by the swept-up mass and the ejecta is
already overrun by the reverse shock. There is also no sign of the Vishniac instabilities \citep{Vishniac83} reported by several numerical studies explaining the observed complex morphology of SNRs \citep[e.g.][]{MacLow93,Miniere18}. These instabilities can develop only if the jump across the shock front is sufficiently high ($\gtrsim$ 10). Shells in the presented simulations do not fulfill this condition, because even though they are relatively thin, the applied resolution does not let them become even thinner and denser and cool below $\sim$10$^{4}$ K. The lack of these dynamical instabilities \citep[see e.g.][]{Moranchel21} is not expected to affect the large-scale picture of how SNe can move gas around near a galactic nucleus, as their impact is subordinate to the importance of tidal forces and the conservation of momentum.

In order to compare the evolution of the shell in the \textsc{flash} and \textsc{ring} simulations, we introduce the equivalent radius, which is the radius of a sphere containing, for a given ambient density,  the same mass as the SN remnant in the simulations. The equivalent radii are plotted in Fig. \ref{fig:sn_shell_radius-N3-N4-N5}. 
Starting at the time of shell formation, which is given in
Table \ref{tab:initial_parameters}, the evolution of equivalent radii, during the time when the shell expansion is supersonic, is almost identical between the two simulation codes. At later times, when the expansion velocity of the shell element drops below the local speed of sound, 
the mass accumulation in the \textsc{ring} simulations ceases and the equivalent radius does not grow any more. That happens after $t_\rmn{sub}$ = 25, 65 and 125 kyr in the case of N5, N4 and N3, respectively. In general, the SN shell evolution during the first few tens of kiloyears is in excellent agreement between \textsc{flash}, which solves the hydrodynamic equations on a grid, and \textsc{ring}, which uses thin shell elements that accrete the ISM, regardless of the choice of ambient medium density.

In simulations with \textsc{flash}, as the shell expansion starts to become subsonic, the growth of $M_\rmn{shell}$ slows, but does not stop completely. This is due to the fact that the mass accumulation into the region of increased density continues. Even at low expansion speeds the shell propagates as a sound wave and the encountered mass of increased density is counted as a part of the shell, since there are  cells with $S_\rmn{sh} > 0.5$. This explains the difference of equivalent radii between \textsc{flash} and \textsc{ring} at subsonic times.

As a general conclusion for the homogeneous medium, \textsc{ring} provides a quite realistic description of the evolution of an expanding SN shell.



\subsection{Density $n_\rmn{ISM} = 10^2$ cm$^{-3}$}

In the case of the homogeneous medium of density $n_\rmn{ISM} = 10^2$ cm$^{-3}$ (N2), a slightly different simulation setup is required compared to those described in Sec. \ref{sec:setup_flash}. The lower density causes a more rapid expansion of the shell, thereby engulfing a larger volume during the Sedov-Taylor phase. In the N2 medium, the thin shell is formed later at larger $r_{shell}$ (see Table 1) compared to the higher density models.  Moreover, the faster expansion also requires a larger physical domain in the \textsc{flash} simulation, thus, the computational volume for our \textsc{flash} runs was increased to a full width of 29 pc around the galactic center. Due to the limitation of the computational resources, the larger computational domain inevitably reduces the spatial resolution of the simulation. The physical domain is divided into the same number of cells as in the N3-N5 simulations to keep the computational cost on the same level. Due to the larger computational volume the linear size of an individual cell in the N2 simulations increases to 0.056 pc. The SN explosion site is set at the coordinates x = 10 pc, y = 0, z = 0 to follow the expansion for a longer period before the central region affects the shell in the \textsc{flash} simulation.

\begin{figure}
	\includegraphics[width=0.98\columnwidth]{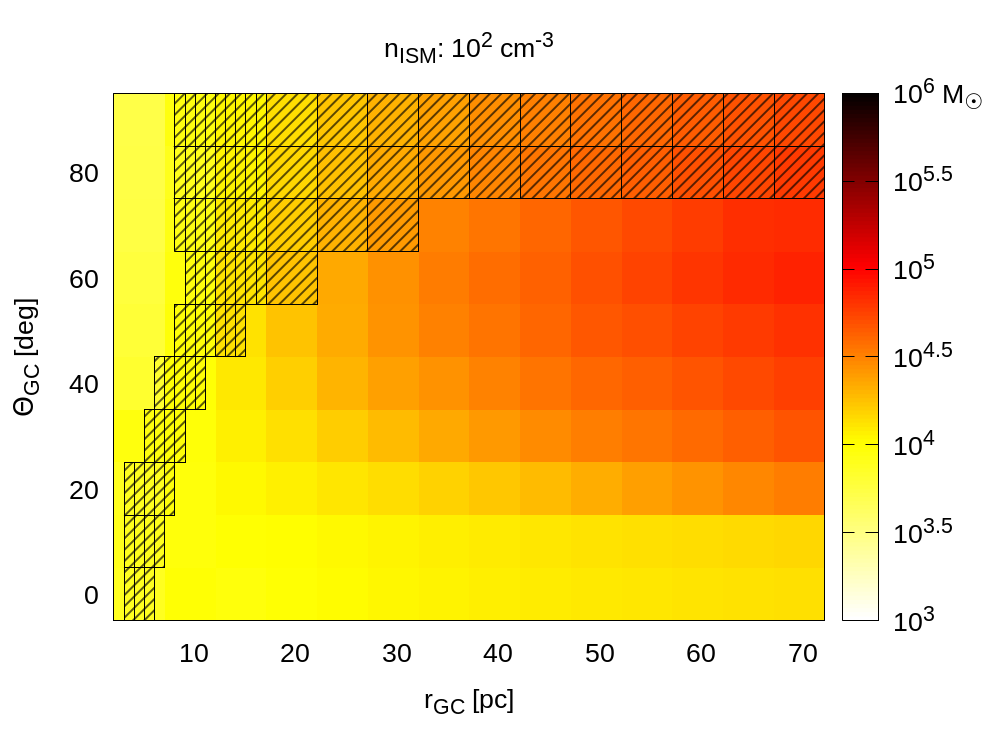}
    \caption{The color shading shows the total mass $m_{sh, tot}$ in a single expanding shell at the end of mass accumulation as a function of the galactocentric distance $r_{GC}$ and the elevation angle $\theta_{GC}$  for $n = 10^2$ cm$^{-3}$. The mass delivery to the central parsec is shown with hatching: the areas without hatching do not deliver any mass, single hatched areas deliver 1 - 10\% of the total shell mass, the double hatched areas deliver more than 10\% $m_{sh, tot}$.}
    \label{feeding}
\end{figure}

The size of the shell at the time of its formation implies that the \textsc{ring} simulation cannot start with a spherically symmetric expansion, since its shape and expansion velocities at the time, when the thin shell is created, are influenced by differential galactic rotation. This is why we start the \textsc{ring} simulations already from a small, but small  radius, which is the same as in the case of \textsc{flash} simulations at the time of shell formation, and insert the $E_{th}$ and $E_{kin}$ (see Table 1) derived by \textsc{flash} for the time of shell formation. We mimic the expansion with the \textsc{ring} thin-shell approximation during the Sedov-Taylor phase before formation of the thin shell, keeping the internal $E_{th}$ constant. At the time of thin shell formation, we start the expansion with velocities that are already influenced by galactic differential rotation, and we start decreasing again the energy density interior to the shell in inverse proportion to its growing volume. 

The column density views are compared in Fig. \ref{fig:sn_shell_evolution_n2}. The size of shell is marginally larger in \textsc{flash}, which is probably due to the effect of slightly different initial conditions related to different resolution in \textsc{flash} and \textsc{ring}. At later times, the forward "beak" of the shell becomes elongated compared to that in \textsc{ring} as it reaches the central region. In the \textsc{flash} simulations, the medium is not entirely homogeneous inside the central 1 pc, due in part to the imposed discontinuity in the rotational velocity at 0.5 pc distance from the center. The inhomogeneities generate turbulence and the related turbulent viscosity leads to angular momentum loss and faster inward migration of gas toward the center compared to \textsc{ring}, where the ISM is always homogeneous.

\begin{center}
\begin{table}[hbt]  
\centering
\begin{tabular}{ l r r r r}
$r_{GC}$ & $n_\rmn{ISM}$ & $\overline{{m^{x}}_{feed}}$ & ${f^{x}}_{dlvr}$ & $V_{dlvr}$\\
 $[pc]$ &  $[cm^{-3}]$ & $[M_{\odot}]$ & $[\times 10^{-3}]$ & $[10^3\ pc^3]$ \\
\hline
 25  & $10^2$ &  62 & 341 & 22 \\ 
     & $10^3$ & 197 & 174 & 11 \\
     & $10^4$ & 409 &  72 & 4.7 \\
     & $10^5$ & 551 &  46 & 3.0 \\
      \hline
 50 & $10^2$ &  50 & 124 & 65 \\ 
    & $10^3$ & 160 &  74 & 39 \\
    & $10^4$ & 150 &  32 & 17 \\
    & $10^5$ &  69 &  5.8 & 3.0 \\ 
 \hline
 100 & $10^2$ &   34 & 46 & 193 \\ 
     & $10^3$ &  120 & 23 & 94 \\ 
     & $10^4$ &   21 &  5.1  & 22 \\ 
     & $10^5$ &    9 &  0.71 & 3.0 \\ 
 \hline
 200 & $10^2$ &  48 & 19 & 637 \\ 
     & $10^3$ &  28 &  6.7 & 220 \\ 
     & $10^4$ &   3 & 0.64  & 22 \\    
     & $10^5$ &   1 & 0.085  & 2.9 \\        
\hline
240 & $10^2$ & 60 & 17 & 1000 \\
    & $10^3$ & 16 & 3.9 & 224 \\
    \hline
400 & $10^2$ & 80 & 6.6 & 1780 \\
 & $10^3$ & 3.5 & 0.84 & 224 \\
\hline
\end{tabular}
\caption{\label{N2-N3-N4-N5-results}Results for the 4 density cases as a function of the
region within $r_{GC}$ of the galactic center that was explored. ${m^x}_{feed}$ is the average mass fed to the SMBH per supernova
exploding within the radius considered. ${f^x}_{dlvr}$ is the ratio between the volume of the region from where SN can deliver the mass to the central 1 pc to the total volume of the  studied sphere. ${V_x}_{dlvr}$ is the volume of the region from where the mass can be delivered to the SMBH, it is the result of the network of \textsc{ring} simulations.}
\end{table}
\end{center}

\begin{figure*}
	\includegraphics[width=17cm]{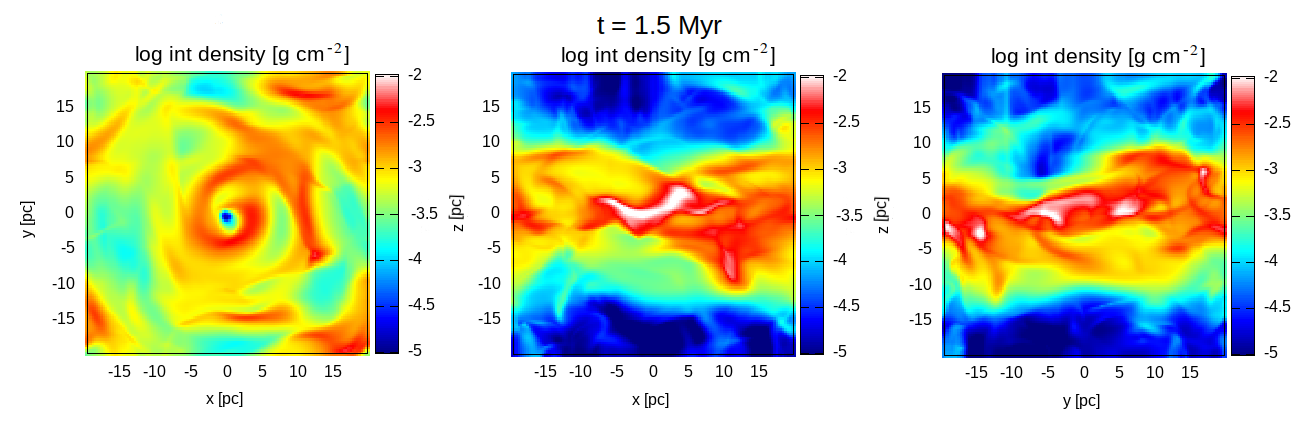}
    \caption{Column densities from three directions of the turbulent 3D "initial" distribution  at the moment before the SN explosion. }
    \label{fig:turbox16_disc}
\end{figure*}

The equivalent shell radii derived in the case of N2 from \textsc{flash} and \textsc{ring} simulations are close to each other, as they are in the  N3, N4, and N5 models (see Fig. \ref{fig:sn_shell_radius-N2}). 
The small discrepancy between the two codes even  before $t_\rmn{sub}$ can be attributed to slightly different initial conditions originating from mapping the \textsc{flash} 3D shell to the infinitesimally thin layer of \textsc{ring}, which depends on resolution. In the case of \textsc{flash} simulations, the increased densities in front of the shell are also included in the mass of the shell, elevating its mass above the shell mass computed by  \textsc{ring}. Increasing the resolution leads to a reduction of the density enhancement in front of the shell, which brings the equivalent radii derived from \textsc{flash} closer to that of \textsc{ring}. Moreover, minor differences may originate from the less efficient cooling of the lower spatial resolution \textsc{flash} simulations, where the energy loss is lower due to the lower density peak of the shell. 

\subsection{Shells expanding into homogeneous density media: N2, N3, N4, N5 \textsc{ring} models.}
\label{sec:uniform}
Now, in order to compare with the results of Paper I, we again include in RING the part of the gravity force perpendicular to the rotational plane.
In Figure \ref{feeding} we show for the N2 case the total mass $m_{sh, tot}$ collected within a single SN shell at the end of mass accumulation when all the shell becomes subsonic. It complements Fig. 8 of  Paper I, which shows the N3, N4 and N5 cases. The initial SN positions for which high total masses can be contributed  to the central parsec. $m_{sh, tot}$ shift in the N2 model to larger galactocentric distances $r_{GC}$ beyond 70 pc from the galactic center and they lie close to the rotational axis. But the mass delivered into the central parsec is less even when the volume of the region from where the supernovae can deliver the mass into the central 1 pc is larger.


\begin{figure*}
    \centering
     \includegraphics[width=15cm]{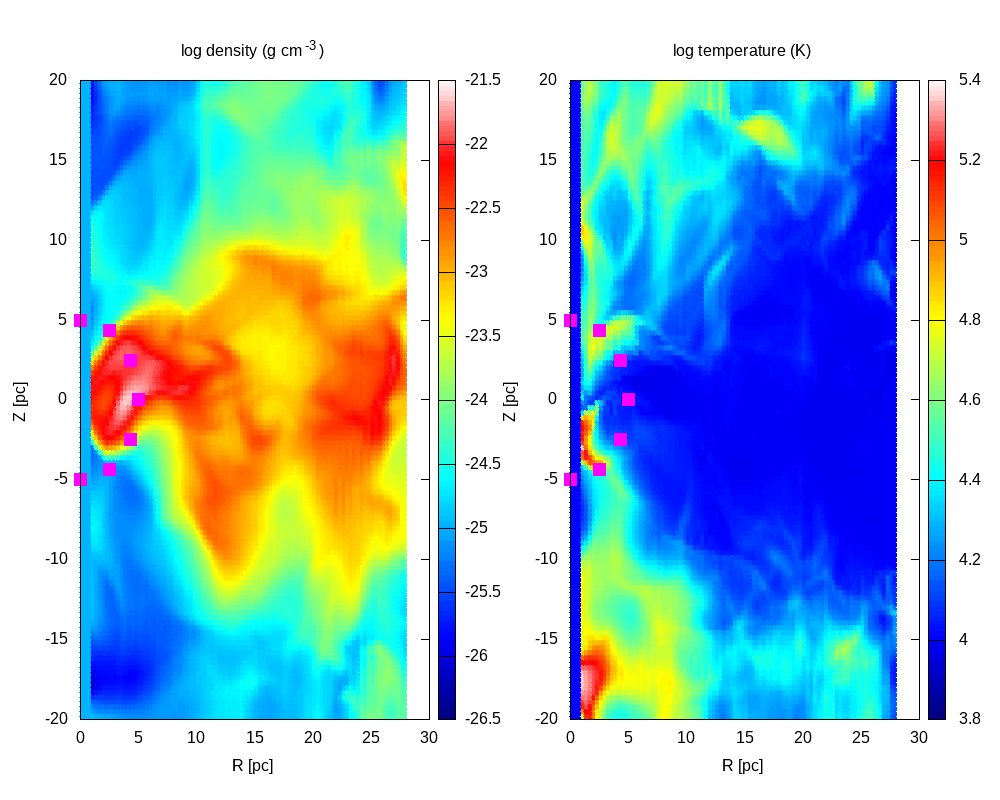}
    \caption{ISM distribution from \textsc{flash} simulations, which serve as the input for RING
      simulations (left: density, right: temperature), represented with axially averaged maps. Purple squares are the chosen positions of the individual supernova explosions.}
    \label{3D-axav}
\end{figure*}

\begin{figure*}
    \centering
	\includegraphics[width=16cm]{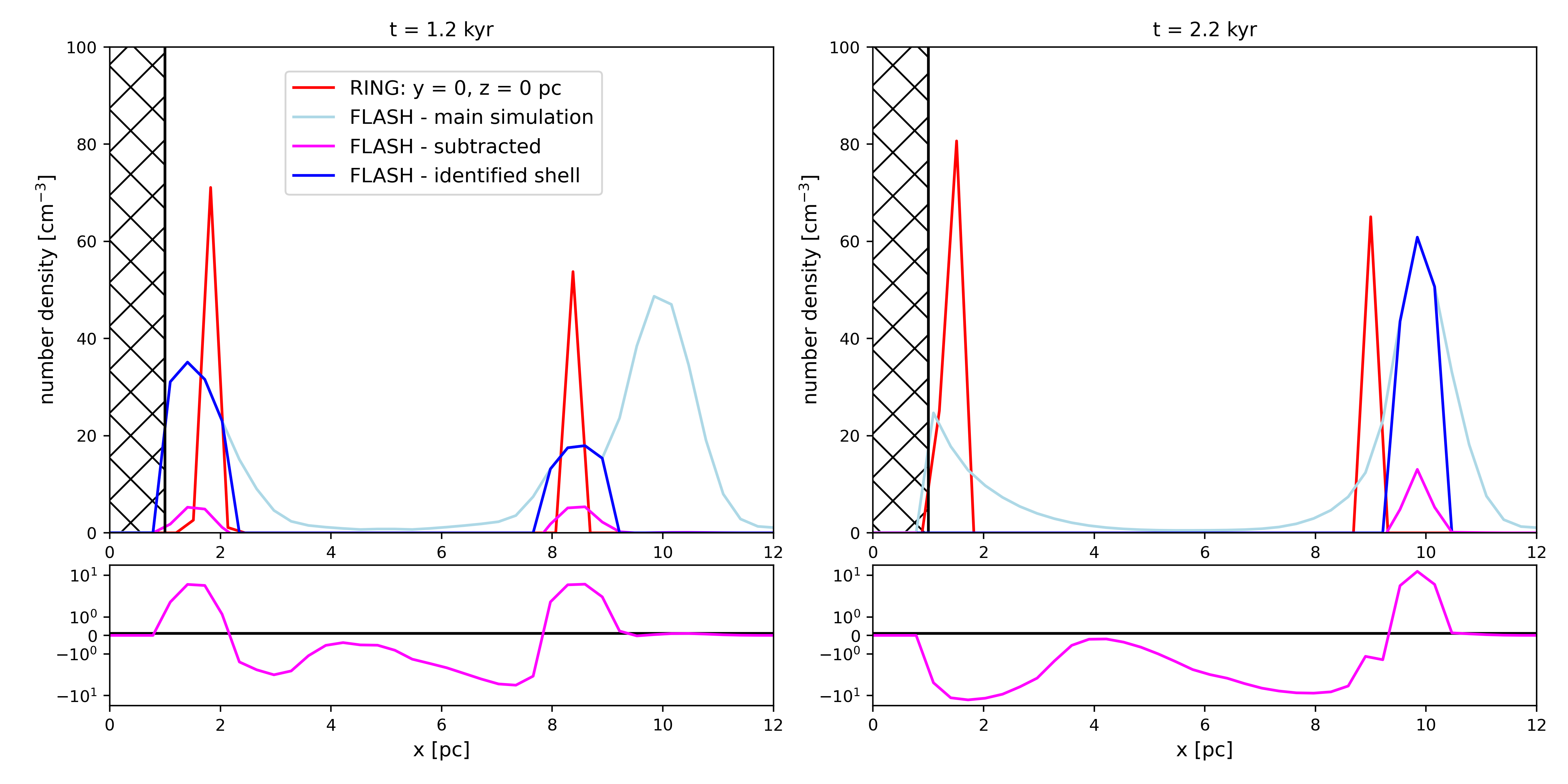}
    \caption{\textbf{Number densities along the x-axis by \textsc{ring} and \textsc{flash} in  turbulent box simulations with expansion starting at (x, y, z) = (5, 0, 0) pc. In the case of \textsc{ring}, the infinitesimally thin shell is resampled with the resolution of 0.31 pc. For \textsc{flash}, the remaining density after the subtraction of the reference simulation from the main simulation is shown by the magenta lines, and the total density at the places of the SNR shell as shown by the magenta lines are given by dark blue lines. The hatched region illustrates the innermost 1 pc, where the inwardly moving is deposited by both the main and the reference simulations. Lower panels show the density after the subtraction of the reference simulation with the expanded number density scale.}}
    \label{fig:ring-flash-1D}
\end{figure*}


\begin{figure*}
    \centering
	\includegraphics[width=16cm]{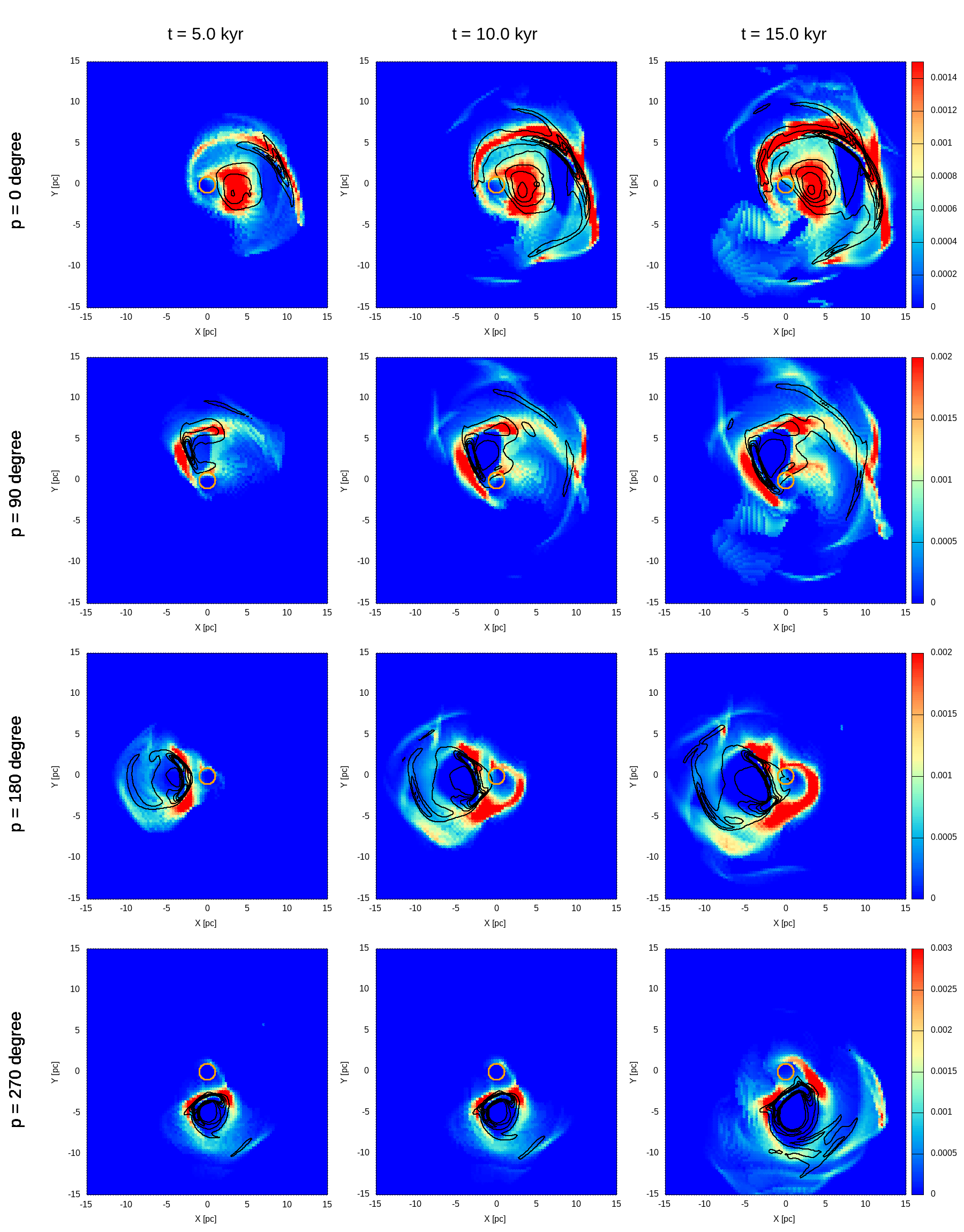}
    \caption{Column density maps of the shell evolution in the \textsc{ring} (contours) and \textsc{flash} (colour maps)  simulations. The four rows illustrate the effects of four different explosion sites in the xy rotational plane, at r$_{GC} = 5$ pc around the GC. \textbf{The five levels of isocontours cover the same column density range as the colorbars in each row.}}
    \label{ring-flash-shapes}
\end{figure*}

The results of a network of \textsc{ring} models for the various ambient ISM densities are given in Table \ref{N2-N3-N4-N5-results}. For SN locations closest to the SMBH (e.g. the case of $r_{GC}= 25\ pc$) high-density cases bring more mass to the SMBH. On
the other hand, for more distant SNe (e.g. the case of $r_{GC}= 200\ pc$), low-density cases are more effective at delivering matter to the SMBH, since SNe exploding in such a medium have relatively large remnants, and have a much higher chance
to reach the SMBH vicinity even if they originate at large distances. 
The size of the region from where the SN delivers the mass into the central 1 pc depends on density: the smaller is the density, the larger is the region.. In the case we study smaller galactocentric distances, we cover only parts of  the region from where mass can flow to the center.

The mass accumulated in the shell elements is influenced by their motion relative to the interstellar medium. This is a combination of shell expansion, galactic rotation and motion under the influence of the force perpendicular to the galactic plane not balanced by rotation, i.e. the $K_z$ force. In Table \ref{N2-N3-N4-N5-results}, Figure \ref{feeding} and in Figure 8 of Paper I,  the average shell mass and the original SN locations from where a fraction of the expanding shell can feed the central parsec are affected by all three velocity contributions. According to our preliminary results, the $K_z$ force influences mainly the feeding from large galactocentric distances above the plane of rotation.  In a subsequent paper we shall disentangle individual contributions in the case of different homogeneous media. In the remainder of this paper we focus on inhomogeneous density distributions, where the density at high z distances is reduced, which also decreases the influence of the $K_z$ force on the final feeding rate.


\section{Inhomogeneous medium}
\label{turbulent}

\subsection{Imposed turbulence}

In order to mimic more realistic ISM density distributions, we also compare \textsc{ring} and \textsc{flash} simulations in media that are inhomogeneous in both density and velocity. For \textsc{flash}, we adopted a time-varying force field pushing the gas into turbulent motion modifying  the rotation described in Sec. \ref{sec:grav_rot}. We adopt a commonly used procedure, implemented in the {\it Stir} module of the \textsc{flash} code. It randomly generates Fourier modes of the force field characterized by wavevector $\mathbf{k} =  2\pi\mathbf{m} / L$, where $L$ is the size of the computational domain and $\mathbf{m}$ is the direction vector with integer components, and evolves them using the generalized Ornstein-Uhlenbeck process \citep[see][for details]{Eswaran88, Schmidt09}. We let the algorithm generate modes in range $3 < |\mathbf{m}| < 7$, with the amplitudes experimentally chosen to form a thick disk with the root-mean-square velocity of turbulent motion close to $\sim 30$\,km/s. This value was set to maintaining the disk thickness close to $10$\,pc.
In a separate, future communication, we shall explore media with different levels of turbulence driving.

\begin{figure*}
	\includegraphics[width=17cm]{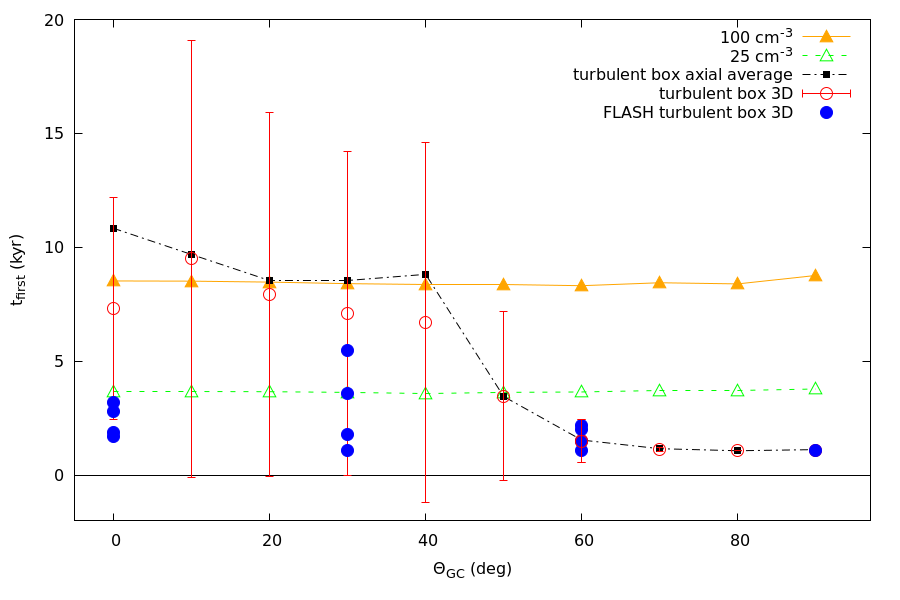}
    \caption{Times of the first touch of the SNR with the central 1 pc sphere, for a SN starting at r$_{GC}$ = 5 pc. The open or solid triangles give the results of the \textsc{ring} simulations in a homogeneous medium of density 25 or 100 particles per cm$^{-3}$, open circles give the \textsc{ring} results in a turbulent 3D medium as is shown in Fig. \ref{fig:turbox16_disc}, solid squares give the \textsc{ring} results in the axially averaged turbulent distribution shown in Fig. \ref{3D-axav}. The first touch times derived from  \textsc{flash} simulations are represented with solid circles. The vertical bars give the rms values of first touch times derived from \textsc{ring} simulations at given $\theta_{GC}$.
    }
    \label{first}
\end{figure*}

The turbulent simulation is evolved to the point when the average density is $n = 10^{2} - 10^{3}$ cm$^{-3}$ within $\pm$5 pc from the galactic plane, but the large-scale motion of the gas is still dominated by the initial velocity profile described in Sec. \ref{sec:grav_rot}. This defines the turbulent 3D "initial" conditions at time $t_\rmn{zero} = 1.5$ Myr, when the SN explosion is inserted. The column density of this turbulent 3D "initial" distribution is shown in Fig.\ \ref{fig:turbox16_disc}.

\begin{figure*}
    \includegraphics[width=17cm]{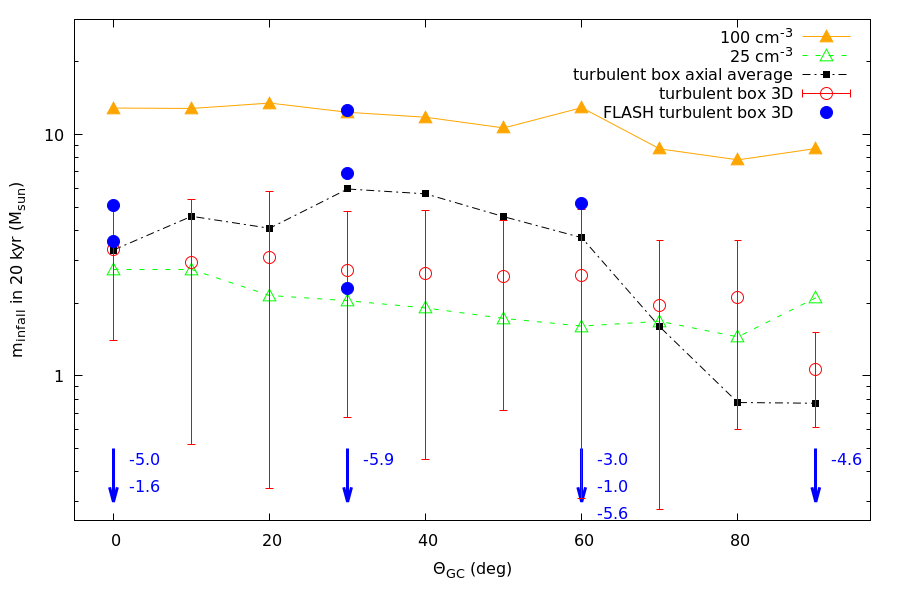}
    \caption{Mass reaching into the central 1 pc as a result of a SN at distance of 5 pc during the first 20 kyr of the SNR expansion in homogeneous and/or turbulent medium. The SNs increase the ISM migration into the central 1 pc, but in some cases the SNRs disturb it  and decrease the migration in comparison to the situation without a SN explosion. This negative cases are shown with arrows and the amount of the migration decrease. The symbols are the same as in Fig. \ref{first}.
    }
    \label{infall}
\end{figure*}

In the following \textsc{flash} simulations the density, temperature, and velocity fields develop together with the expanding SN shells. 
In parallel \textsc{ring} simulations, this "initial" density and temperature are fixed. Velocities follow the rotation curve described in Section \ref{sec:grav_rot} (see equation \ref{eq:quadratic}). Although \textsc{ring} uses the fixed 3D initial density from \textsc{flash}, 
and the evolution of the turbulence is not followed,  the impact of possible density changes within the turbulent box  on the short $\sim 20$ kyr timescale is small.
In \textsc{ring}, we also use the turbulent, axially averaged distribution, which is shown in Fig. \ref{3D-axav}. We see that a turbulent disk is formed, slightly inclined by about 10 - 20 degrees, and with density decreasing and temperature increasing with $z$. To see possible deviations due to changing position angle $p$,  we show in Figs. \ref{first} and \ref{infall} the results of \textsc{ring} simulations in an axially averaged turbulent distribution compared to experiments starting at different position angles $p$ in 3D initial density distribution.

\begin{figure*}
	\includegraphics[width=17cm]{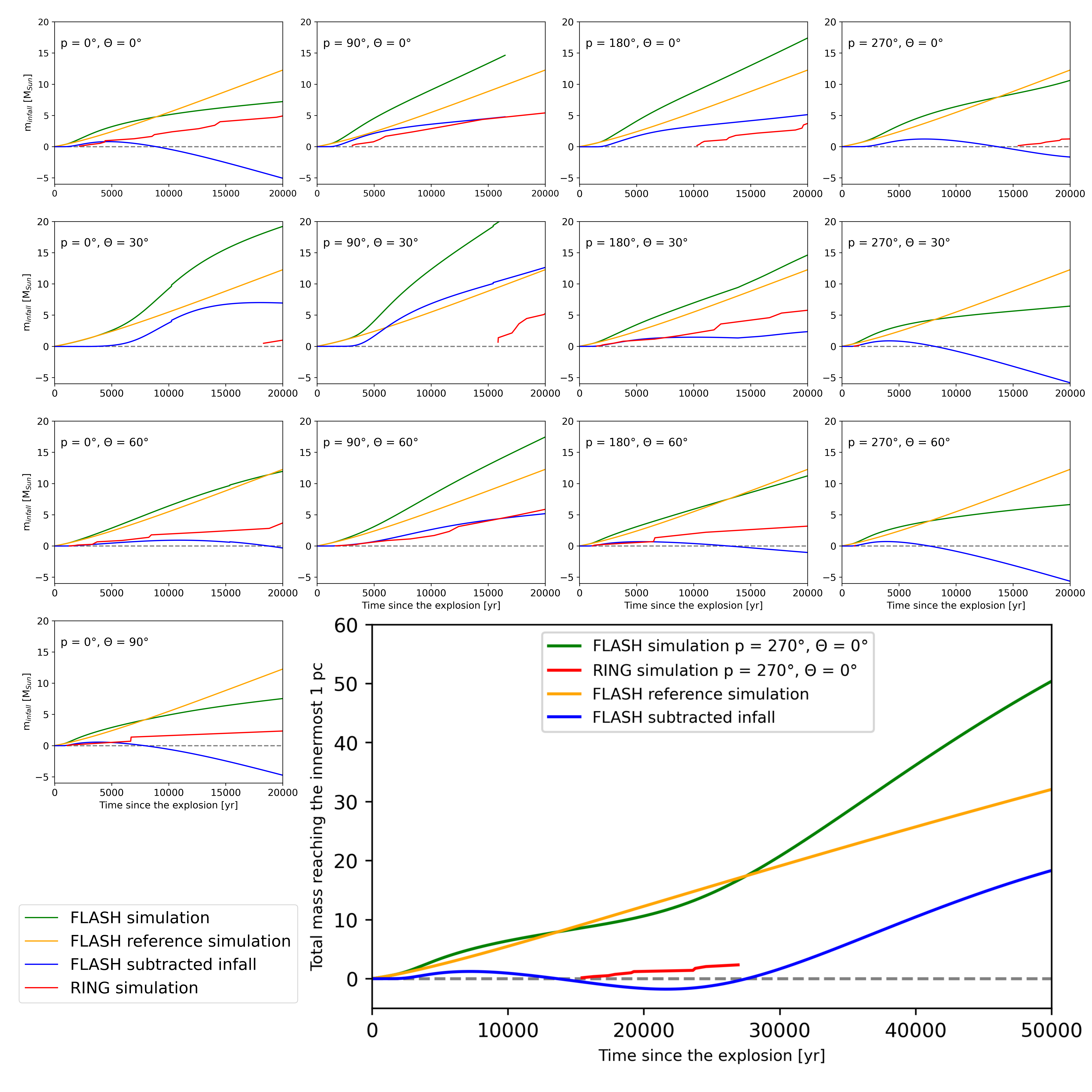}
    \caption{The amount of gas reaching the innermost 1 pc in \textsc{ring} and \textsc{flash} simulations. For \textsc{flash}, the estimate the infall rate due to the impact of the expanding shell, the infall function of a reference simulation ($m_\rmn{inf,ref}(t)$, orange) is subtracted from the registered infalling gas of the simulation ($m_\rmn{inf}(t)$, green). Where the difference of these function leads to a decreasing trend, a constant value is subtracted from $m_\rmn{inf}(t)$ (see text).}
    \label{fig:infall_mass_flash}
\end{figure*}

\subsection{SN explosions at $r_{GC} = 5$ pc: first 20 kyr and beyond}

To compare the \textsc{flash} and \textsc{ring} simulations of the evolution of SN shells in different regions of the turbulent medium, we simulate explosions at various azimuthal positions and elevation angles, all located at $r_{GC} = 5$ pc (see magenta squares in Fig. \ref{3D-axav}.
\textbf{However, the detection of the SNR shell is more difficult than in the case of the homogeneous medium, because the Mach-number criterion described in Sec. \ref{sec:setup_flash} cannot be used due to the highly turbulent velocity field. Moreover, local density clumps produced by the turbulent field may exceed the shell density (see the upper left panel of Fig. \ref{fig:ring-flash-1D}, where the light blue line shows the peak density at about 10 pc in the FLASH simulation, while the shell formed in RING simulations is identified at $\sim$ 8 pc)}.

\textbf{To check the impact of the shell, a reference simulation is created with exactly the same properties, but without the SN explosion (giving $\rho_{\rmn{ref}}$). The 3D density map of the reference simulation is subtracted from the main simulation ($\rho_{\rmn{main}}$, light blue colour in Fig. \ref{fig:ring-flash-1D}), showing the parts of the shell where the mass collected by the expanding shell exceeds the density of the background ($\rho_{\rmn{res}}$). The residual is shown by magenta lines in Fig. \ref{fig:ring-flash-1D} that mark the derived location of the SNR shell. The dark blue line in Fig. \ref{fig:ring-flash-1D} gives the total density ($\rho_{\rmn{SNR}}$) at places where the shell is found to surpass the threshold given below. This mass is considered as a part of the shell:}

\begin{equation}
\rho_{\rmn{SNR}} = 
\begin{cases}
\, \rho_{\rmn{main}}
& \text{where  } \rho_{\rmn{res}} \geq 0.1 \, cm^{-3} \\
\\
\, \, \, \, 0.0
& \text{where  } \rho_{\rmn{res}} < 0.1 \, cm^{-3}
\end{cases}
\label{eq:rho_snr}
\end{equation}

\noindent
\textbf{where $\rho_{\rmn{res}} = \rho_{\rmn{main}} - \rho_{\rmn{ref}}$ is the difference of the density maps of the main and the reference simulations. To exclude the small fluctuations in the residual density map, a rather low threshold was defined.}

\textbf{Note that this definition of the shell is an operational one meant to highlight those areas of the simulation where extra mass was concentrated by the shock. The residual density map includes not only positive and near-zero, but also negative values, from where the shell has excavated some material as it has expanded. As our examples show, this indirect identification of the shell works well for the locations affected by the shock. At the same time, any kind of precise mass estimation of the shell is problematic for various reasons, e.g., the ambiguous density limits of the shell (see right panel of Fig. \ref{fig:ring-flash-1D}) or a possibly insufficient value of the chosen threshold in Eq. \ref{eq:rho_snr}.} 

\textbf{Since a direct quantitative comparison between \textsc{flash} and \textsc{ring} shell masses through the evolution of the shell is impractical,} we have relied on visual inspection to compare the time evolution of the shell between the two simulations (see Fig. \ref{ring-flash-shapes}). 
We see that, in \textsc{flash}, the shell always expands slightly further than in \textsc{ring} in a given time interval, however, the agreement is very good in general: the overall shape and spatial extent of all the shells are comparable, high-density clumps are formed at similar positions, and the same low-density extensions appear within the two different simulations. \textbf{The match between the location of the shells is typically better than the thickness of the shell wall of \textsc{flash}.}
Most of the differences between \textsc{flash} and \textsc{ring} are due to the fact that \textsc{ring} does not capture naturally-occurring instabilities in the walls of the expanding shell and that the \textsc{ring} simulations expand to non-evolving ISM distribution. The instabilities that are seen in the simulations with \textsc{flash} form  holes that channel the hot gas from the bubble interior to the outside of the shell, thereby increasing the affected volume.

In Fig. \ref{first}, we display the time of the first touch for the various simulations, defined as the moment when some portion of the shell penetrates the central 1 pc radius sphere. The times derived from \textsc{flash} are complemented with more experiments with \textsc{ring} starting from additional positions on the $r_{GC} = 5$ pc sphere. We show the time of first touch and its root mean square dispersion based on all the experiments performed at different position angles, {\it p} (it is the angle between SN explosion and the galactic center projected to the $(X, Y)$ plane measured anticlockwise from the positive direction of the $X$ axis), with the \textsc{ring} code applied to a turbulent 3D density distribution at the given elevation angle, $\theta_{GC}$, and we compare to the first touch times in homogeneous density distributions of 100 and 25 particles per cm$^{-3}$, as well as to experiments performed in a turbulent density distribution that is axially averaged. In \textsc{flash} the first touch happens typically earlier than in \textsc{ring}, because the instabilities followed by \textsc{flash} open a faster track towards the central 1 pc. However, the \textsc{flash} values still lie within the rms dispersion of the \textsc{ring} simulations in the majority of cases.  
We see that the time of first touch in the inhomogeneous models is highly dependent on the elevation angle: for $\theta_{GC} > 50$ degrees the first touch occurs well before 20 kyrs and the spread of the first touch times is rather small, while for $\theta_{GC} < 50$ degrees the spread of the first touch times is large, from some position angles $p$  it is more than 20 kyrs.  This is due to the turbulent nature of the ISM, where the shell encounters low- and high-density places. The density spread is much larger at low $\theta_{GC}$ values than at large $\theta_{GC}$ explaining the changing spread in first touch times for different $\theta_{GC}$.

As an approximate quantitative comparison, we also estimate the mass of the gas migrating into the central 1 pc. The results for the \textsc{ring} simulations are shown in Fig. \ref{infall} for the mass delivery during the first 20 kyrs.  We see that the mass delivery in the turbulent medium is well mimicked by the homogeneous medium with 25 particles per cm$^{-3}$, which is close to the average density observed in the turbulent medium  within 5 pc.     
In the case of the \textsc{flash} simulation, the infalling mass $m_\rmn{inf}(t)$ cannot be estimated directly, as some gas would migrate into the central parsec even without the direct influence of the expanding SN shell, due to the turbulent viscosity in the ISM. Therefore, the inwardly migrating mass of the reference simulation, $m_\rmn{inf,ref}(t)$, is subtracted from that of the SNR simulation to derive the net effect of the SN on the mass feeding. This procedure assumes that, besides the presence of the expanding shell, the SN explosion does not influence the motion of the ISM. In some cases $m_\rmn{inf}(t)$ actually declines below $m_\rmn{inf,ref}(t)$, since the supernova can blow away material that would otherwise migrate into the 1 pc central sphere. On average, however, the net effect of a nearby SN is to enhance the amount of mass fed to the central parsec.

A variety of different situations is shown in  Fig. \ref{fig:infall_mass_flash}, where we give the integrated mass that migrates into the central 1 pc from different initial positions on the $r_{GC}=5$ pc sphere as computed by \textsc{flash} and \textsc{ring}. It reflects a complicated density and velocity landscape leading sometimes to a good \textsc{flash}-\textsc{ring} agreement in size and shape of the SNR, but also in some situation to a disagreement in the amount of mass delivered to the central 1 pc. As visible in Fig. \ref{infall}, in the positive cases, the migrated masses are in good agreement what suggests that fast code \textsc{ring} can be used for statistical studies even in strongly turbulent medium. In one of the negative cases, ($\theta_{GC} = 0$, {\it p} = 270 degrees), we show the subsequent evolution up to 50 kyrs. The early decline in migration towards the center is later reversed to increased mass migration connected to dense layers of the SNR arriving to the central region. The mass collected in the central 1 pc by migration should be estimated with a more statistical approach, which will be performed in the future.

\section{Conclusions}
\label{conclusions}


In this study, we present a comparison between two simulation codes for the the 3D evolution of expanding SN shells in the vicinity of the Galactic Center. The first one is the simplified code \textsc{ring}, which uses the thin SN shell approximation, and the second is \textsc{flash}, which is a more physically complete but computationally expensive hydrodynamic code. The aim is to demonstrate the viability of using relatively fast \textsc{ring} simulations for mapping the broad parameter space of initial and boundary conditions.

In homogeneous media having particle densities in the range $25 - 10^5$  cm$^{-3}$, there is compelling agreement between \textsc{flash} and \textsc{ring} in shapes and equivalent radii throughout the whole shell evolution until it becomes subsonic. In  highly inhomogeneous, turbulent media, the differences can be explained mostly as the effect of insufficient resolution and slight differences in the initial conditions.  There is good agreement of shapes and sizes of shells, and times of the first touches of the central 1 pc sphere. The delivered masses deviate in some cases, however, both simulations still predict the same order of magnitude of the delivered mass.
In the case of inhomogeneous media, additional factors account for the differences between \textsc{ring} and \textsc{flash} results, e.g. different turbulent driving mechanisms, dynamical evolution of the ISM, and instabilities.
We conclude that \textsc{flash} and \textsc{ring} simulations both yield the results that for supernovae occurring at a distance of 5 pc from the galactic center, the mass delivered within 20 kyr by one SN is typically about 3 M$_\odot$. 

In order to better quantify the mass delivery to the central parsec, we need to statistically investigate the impact of randomly distributed SNe occurring throughout larger volumes, with $r_{GC}$ up to $\sim 25$ pc. In a later publication, we plan to adopt an approach similar to that which we employed in Paper I for  homogeneous media, and apply it to the case of more realistic, inhomogeneous, turbulent media. According to our results, \textsc{ring} provides a fast but reliable tool for future statistical studies.

\section*{Acknowledgements}

The authors acknowledge the support by the Czech Science Foundation Grant 19-15480S and by the project RVO: 679856815. This work was supported by The Ministry of Education, Youth and Sports from the Large Infrastructures for Research, Experimental Development and Innovations project ‘IT4Innovations National Supercomputing Center - LM2015070’. BB received support from the Hungarian NKFIH/OTKA FK-134432 grant.
The software used in this work was developed in part by the DOE NNSA ASC- and DOE Office of Science ASCR-supported Flash Center for Computational Science at the University of Chicago.

\section*{Data Availability}

The data underlying this article will be shared on reasonable request to the corresponding author.










\bsp	
\label{lastpage}
\end{document}